\documentclass[twocolumn,citeautoscript,showpacs,preprintnumbers,amsmath,amssymb,prl,floatfix,footinbib]{revtex4}
\usepackage{graphicx}
\usepackage{dcolumn}
\usepackage{bm}
\usepackage{times}
\begin{document}
\title{Magnetic ordering in EuRh$_2$As$_2$ studied by x-ray resonant magnetic scattering}
\author{S.~Nandi$^{1}$, A.~Kreyssig$^{1}$, Y.~Lee$^{1}$, Yogesh~Singh$^{1}$, J.~W.~Kim$^{2}$, D.~C.~Johnston$^{1}$,
B.~N.~Harmon$^{1}$, and A.~I.~Goldman$^{1}$}
\affiliation{\\$^{1}$Ames Laboratory, US DOE and Department of
Physics and Astronomy, Iowa State University,
Ames, Iowa 50011, USA\\
$^{2}$Advanced Photon Source, Argonne National Laboratory, Argonne,
Illinois 60439, USA}

\date{\today}

\begin{abstract}
Element-specific x-ray resonant magnetic scattering (XRMS)
investigations were performed to determine the magnetic structure of
Eu in EuRh$_2$As$_2$ with the ThCr$_2$Si$_2$ structure. In the
temperature range from 46~K down to the lowest achievable
temperature of 6~K, an incommensurate antiferromagnetic (ICM)
structure with a temperature dependent propagation vector
\boldmath{${\tau}$}~$\approx$~(0~0~0.9) coexists with a commensurate
antiferromagnetic (CM) structure. Angular-dependent measurements of
the magnetic intensity indicate that the magnetic moments lie in the
tetragonal basal plane and are ferromagnetically aligned within the
\textbf{a-b} plane for both magnetic structures. The ICM structure
is most likely a spiral-like magnetic structure with a turn angle of
$\sim$162$^\circ$ (0.9$\pi$) between adjacent Eu planes in the
\textbf{c} direction. In the CM structure, this angle is
180$^\circ$. These results are consistent with band-structure
calculations which indicate a strong sensitivity of the magnetic
configuration on the Eu valence.
\end{abstract}


\maketitle The complex interplay between superconductivity,
magnetism and structural instabilities in $A$Fe$_2$As$_2$ ($A$~=~Ba,
Sr, Ca, and Eu) pnictides upon chemical substitution, or under
applied pressure has generated a great deal of recent attention and
research activity in this ThCr$_2$Si$_2$-type class of
compounds.\cite{rotter_08,ni_08,sasmal_08,alireza_08,torikachvili_08,kreyssig_08,goldman_09}
Related isostructural compounds such as BaRh$_2$As$_2$,
BaMn$_2$As$_2$ and
EuRh$_2$As$_2$\cite{singh:104512,singh_08,singh_08_1} have also been
synthesized and studied in an attempt to significantly increase
superconducting transition temperatures and to understand the role
of Fe in the Fe-As layers as well as the \textit{A} site in
promoting superconductivity. The metallic compound EuRh$_2$As$_2$
shows a plethora of interesting physical properties such as giant
magnetoresistance and a strong reduction in the electronic specific
heat coefficient with applied field in the antiferromagnetic state
below $T_{N}$~=~46~K.\cite{singh_08} The reported intermediate
valence of 2.13(2) for Eu\cite{singh_08} adds further interest as it
may be tuned by applied pressure or
temperature.\cite{rohler_82,chefki_98} Since Eu$^{2+}$ ions carry a
magnetic moment with spin \textit{S}~=~7/2 while Eu$^{3+}$ does not
have a permanent magnetic moment, EuRh$_2$As$_2$ may be an excellent
model system to study the complex interplay between valence and
magnetism in the ``122" pnictides.

EuRh$_2$As$_2$ crystallizes in the ThCr$_2$Si$_2$-type tetragonal
structure with space group $I4/mmm$,\cite{hellmann_2007} and lattice
parameters \textit{a}~=~4.075~{\AA} and \textit{c}~=~11.295~{\AA} at
\textit{T}~=~298~K.\cite{singh_08} Specific heat and magnetization
data indicate that the Eu moments order antiferromagnetically below
$T_{N}$~=~46~K.\cite{singh_08} The magnetic susceptibility parallel
to the \textbf{c} axis increases below $T_{N}$, and decreases when
measured perpendicular to the \textbf{c} axis.\cite{singh_08} This
indicates that the moments are primarily aligned in the basal
\textbf{a-b} plane. Nevertheless, the microscopic details of the
magnetic structure are, as yet, unknown. Here, we report on the
magnetic ordering of Eu moments in EuRh$_2$As$_2$ studied by x-ray
resonant magnetic scattering (XRMS). For the present measurements,
this technique offers several advantages, primarily due to the large
neutron absorption cross-section of the natural isotope $^{152}$Eu.
Moreover, the elemental specificity and superior wave-vector
resolution provided by x-rays can be employed to precisely determine
the magnetic propagation vector, particularly for systems with more
than one propagation vector.

Single crystals of EuRh$_2$As$_2$ were grown using a Pb
flux.\cite{singh_08} For the XRMS measurements, an as-grown
plate-like single crystal with a surface perpendicular to the
\textbf{c} axis and of approximate dimensions
1\,$\times$\,1\,$\times$\,0.1\,mm$^3$ was selected. The sample shows
very similar magnetic behavior to that previously
reported.\cite{singh_08} The XRMS experiment was performed on the
6ID-B beamline at the Advanced Photon Source at the Eu
$L_{\mathrm{I\!I}}$ absorption edge (\textit{E} = 7.611~keV). The
incident radiation was linearly polarized perpendicular to the
vertical scattering plane ($\sigma$-polarized) with a spatial
cross-section of 1.0~mm (horizontal) $\times$ 0.25~mm (vertical). In
this configuration, resonant magnetic scattering rotates the plane
of linear polarization into the scattering plane
($\pi$-polarization). In contrast, charge scattering does not change
the polarization of the scattered photons ($\sigma$-$\sigma$
scattering). Pyrolytic graphite PG~(0~0~6) was used as a
polarization and energy analyzer to suppress the charge and
fluorescence background relative to the magnetic scattering signal.
The sample was mounted at the end of the cold-finger of a displex
cryogenic refrigerator with the \textbf{a-c} plane coincident with
the scattering plane and was measured at temperatures between 6~K
and 50~K.
\begin{figure}
\begin{center}
\includegraphics[clip, width=.45\textwidth]{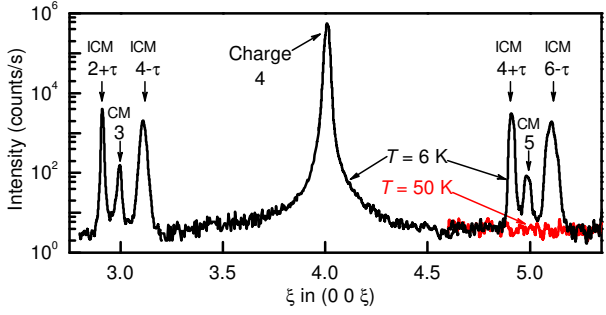}\\
\caption{(Color online) Scan along the (0~0~1) direction at
\textit{T}~=~6~K in the rotated $\sigma$-$\pi$ channel. Note that
the intensity is shown on a logarithmic scale. The data around
(0~0~5) are shown in red at \textit{T}~=~50~K~$>$~$T_N$ for
comparison.} \label{lscan}
\end{center}
\end{figure}

\begin{figure}
\begin{center}
\includegraphics[clip, width=0.4\textwidth]{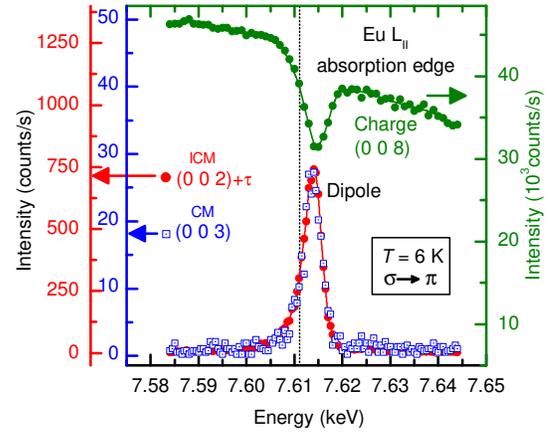}\\
\caption{(Color online) Energy scans of the ICM
(0~0~2)+$\boldsymbol{\tau}$ (red) and CM (0~0~3) (blue) reflections
and of the charge (0~0~8) (green) reflection. The vertical line
depicts the Eu $L_{\mathrm{I\!I}}$ absorption edge as determined
from the inflection point of the charge signal. The solid lines are
guides to the eye.} \label{escan}
\end{center}
\end{figure}

Figure~\ref{lscan} shows a scan along the (0~0~1) direction,
measured at the peak of the dipole resonance (Fig.~\ref{escan}) at
an x-ray energy \textit{E}~=~7.614~keV in the rotated $\sigma$-$\pi$
channel. At \textit{T}~=~6~K~$<$~$T_N$, other than the allowed
charge reflections (0~0~$L$) with $L$~=~even, new satellite peaks
appear which can be indexed as (0~0~$L$)~$\pm$ $\boldsymbol{\tau}$
with $\boldsymbol{\tau}$~$\approx$~ (0~0~0.9), indicating an
incommensurate magnetic structure (ICM). There are also weak peaks
at (0~0~$L$) with $L$~=~odd pointing to an additional commensurate
magnetic structure (CM) with propagation vector (0~0~1)\cite{pv}.
Careful scans along (1~0~0) and (1~1~0) directions reveal no
additional satellite peaks.\cite{scan} To confirm the resonant
magnetic behavior of these peaks, we performed energy scans through
the Eu $L_{\mathrm{I\!I}}$ absorption edge in the $\sigma$-$\pi$
channel (shown in Fig.~\ref{escan}) at 6~K.\cite{atten} We observed
one resonance peak approximately 3.5~eV above the absorption edge
for both ICM and CM structures. This peak arises from dipole
resonant scattering\cite{hannon_88} and confirms that Eu is magnetic
in EuRh$_2$As$_2$.

\begin{figure}
\begin{center}
\includegraphics[clip, width=0.4\textwidth]{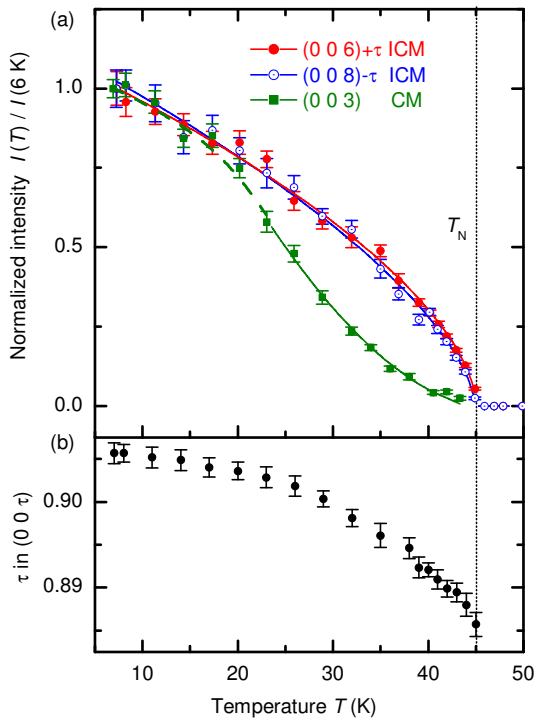}\\
\caption{(Color online) (a) Temperature dependence of the integrated
intensity of the pair of ICM (0~0~6)+$\boldsymbol{\tau}$ (red),
(0~0~8)-$\boldsymbol{\tau}$ (blue) satellite peaks and of the CM
(0~0~3) (green) Bragg reflection, determined by fitting
$\theta$-$2\theta$ scans with a Lorentzian function. The solid lines
are fit to the data as described in the text and the dashed line is
a guide to the eyes. (b) Temperature dependence of the propagation
vector as determined from the pair of ICM reflections.} \label{tdep}
\end{center}
\end{figure}

Figure~\ref{tdep}(a) shows the temperature dependence of the
integrated intensity of the ICM (0~0~6)+$\boldsymbol{\tau}$ and
(0~0~8)-$\boldsymbol{\tau}$ satellite peaks, and the CM (0~0~3)
Bragg reflection. The smooth variation of magnetic intensity close
to the transition temperature indicates that the phase transition is
second order. The integrated intensity ($I\!\sim\!\mu^2$, $\mu$ is
the sublattice magnetization\cite{bruckel_01}) can be fitted with a
power law of the form $I\!\sim\!(1-\frac{T}{T_N})^{2\beta}$. The
obtained exponents $\beta~=~0.32\pm0.02$ and $\beta~=~0.7\pm0.1$ for
the ICM and CM peaks, respectively, will be interpreted later. The
fitted transition temperature, $T_{N}~=~46.0\pm0.5$~K, is in
excellent agreement with the value $T_{N}~=~46\pm$1~K, determined
from the magnetization and heat capacity
measurements.\cite{singh_08} Figure~\ref{tdep}(b) shows the
temperature dependence of the propagation vector after correcting
for the thermal expansion of the lattice. The propagation vector
varies smoothly from 0.905c$^\star$ at 6~K to 0.885c$^\star$ at
46~K, supporting further the incommensurate nature of the ICM
structure.

\begin{figure}
\begin{center}
\includegraphics[clip, width=0.4\textwidth]{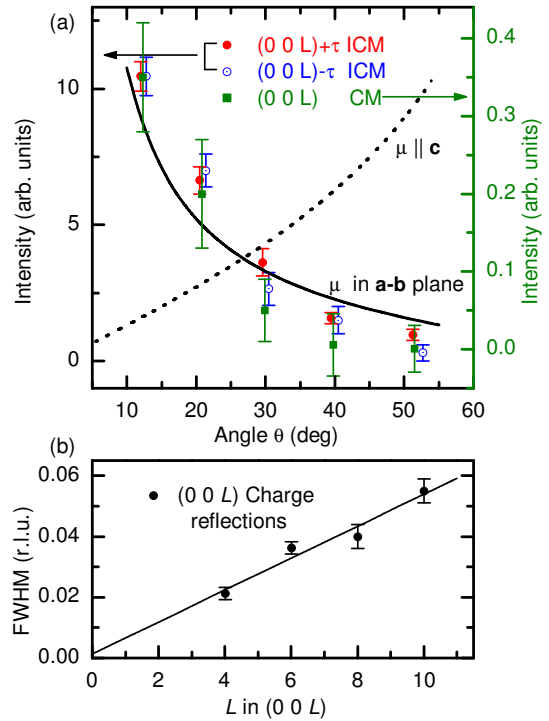}\\
\caption{(Color online) (a) Angular dependence of the integrated
intensity of the ICM (0~0~$L$)+$\boldsymbol{\tau}$ (red),
(0~0~$L$)-$\boldsymbol{\tau}$ (blue) satellite peaks with $L$ being
even and of the CM (0~0~$L$) (green) Bragg reflections with $L$
being odd. The solid and dotted lines represent expected angular
dependence for moments in the \textbf{a-b} plane and in the
\textbf{c} direction, respectively. (b) $L$ dependence of the FWHM
of a series of (0~0~$L$) charge reflections. The solid line
represents the expected $L$ dependence for a sample with variation
in lattice parameter.\cite{dl_note}} \label{qdep}
\end{center}
\end{figure}

We now turn to the determination of the magnetic moment
configuration. For the crystallographic space group $I4/mmm$ and
propagation vectors of the form (0~0~$\tau$), two independent
magnetic representations are possible with moments that are either
strictly along the \textbf{c} direction or confined to the
\textbf{a-b} plane.\cite{wills_00} For a second order phase
transition, Landau theory predicts that only one of the two
above-mentioned representations is realized at the phase
transition.\cite{wills_00} To differentiate between these two
representations, a series of CM and ICM Bragg reflections were
measured. Figure~\ref{qdep}(a) shows the expected angular dependence
of the magnetic intensity for the two above-mentioned
representations along with the observed intensities. The XRMS
intensity for the current experimental configuration can be
calculated as:\cite{hill}
\begin{equation}
\begin{split}
I&=B\frac{(\mu_a\cos\theta)^2}{\sin2\theta},~\textup{for}~\boldsymbol{\mu}
~\textup{in the}~\textbf{a-b}~\textup{plane}\\
&=B\frac{(\mu_c\sin\theta)^2}{\sin2\theta},~\textup{for}~\boldsymbol\mu
\| \textbf{c}
\end{split}
\end{equation}
where $B$ is a scaling factor, $\theta$ is the Bragg angle,
1/$\sin2\theta$ is the Lorentz factor and $\mu_a$ and $\mu_c$ are
the components of magnetic moments along the \textbf{a} and
\textbf{c} directions, respectively. Since the model calculation
with the magnetic moment in the \textbf{a-b} plane closely agrees
with the observed intensity, we conclude that the magnetic moments
lie in the \textbf{a-b} plane for both the ICM and CM structures.

Both a transverse amplitude modulated collinear antiferromagnetic
structure and a basal plane spiral antiferromagnetic structure are
consistent with moments in the \textbf{a-b} plane and a propagation
vector of (0~0~0.9). In an XRMS experiment one cannot distinguish
between these two structures due to the presence of domains.
However, we note that a spiral like structure can persist down to
the lowest temperature whereas a transverse amplitude modulated
magnetic structure must transform to a square-wave modulation due to
the expected equal amplitude of ordered magnetic moments at low
temperatures. Such a ``squaring up" of the magnetic structure would
produce third harmonic satellite peaks $\pm3\boldsymbol{\tau}$ at
\textit{T}~=~6~K, which were not observed (see Fig.~\ref{lscan}).
Therefore, we conclude that the ICM structure is a spiral-like
structure with ferromagnetically coupled moments in each
\textbf{a-b} plane and a temperature dependent turn angle of
$\sim$162$^\circ$ (0.9$\pi$) between adjacent Eu planes. For the CM
structure, the magnetic moments are also ferromagnetically aligned
within the \textbf{a-b} plane. The observation of CM Bragg
reflections at (0~0~$L$) with $L$ odd, together with the absence of
a ferromagnetic signal in magnetization measurements\cite{singh_08}
indicate that the magnetic moments in the adjacent Eu planes are
aligned in opposite directions for the CM structure.

We now turn to the discussion of certain subtle features observed in
the XRMS study. First of all, from Fig.~\ref{lscan}, we note that
the Full-Width-at-Half-Maximum (FWHM) for pairs of satellite
reflections, for example (0~0~4)$\pm$$\boldsymbol{\tau}$, is quite
different and there is an overall increase in FWHM with increasing
scattering angle. Such features in FWHM can be explained assuming a
variation in the lattice parameter $\Delta c\sim$~0.05~{\AA}, and a
related variation in the propagation vector
$\Delta\boldsymbol{\tau}\sim$~0.03\textbf{c}$^\star$. Simultaneous
variations in both \textbf{c} and $\boldsymbol{\tau}$ compensate
each other for the positions of the +$\boldsymbol{\tau}$ satellite
peaks and result in an unchanged FWHM. The effect is opposite for
the -$\boldsymbol{\tau}$ satellites, yields a strong variation in
the positions for the -$\boldsymbol{\tau}$ satellites and,
therefore, increases the FWHM significantly. Indeed, the variation
in lattice parameter and corresponding inhomogeneity in the sample
is also evident from the linear increase of FWHM of different charge
peaks as a function of $L$ [see Fig.~\ref{qdep}(b)] as $\Delta
L$~$\approx$~$\frac{\Delta c}{c}L$ and gives $\Delta
c$~$\sim$~0.05~{\AA}. Here we note that effects other than the
variation in lattice parameter, such as strain in the sample, also
affect the FWHM as a function of scattering angle.

Next, we turn to the observed coexistence of ICM and CM structures
over the investigated temperature range. In rare earth intermetallic
systems a coexistence of CM and ICM structures is
rare,\cite{hill_96, llobet_05} and can arise from minority phases
due to strain, disorder and/or slightly varying stoichiometry in the
sample. The absence of satellite reflections with a combination of
both propagation vectors suggests that the two magnetic structures
are independent. The intensity of the CM peaks is approximately two
orders of magnitude lower than the ICM peaks and indicates a similar
ratio in volume fractions for the CM and ICM phases. In the case of
Eu based rare earth intermetallic compounds such as EuPd$_2$Si$_2$
and EuCu$_2$Si$_2$, a minor phase has been observed which also
orders magnetically at low temperatures with a slightly lower Eu
valence than in the main phase.\cite{abd_85, moritz_87,
bauminger_73}

To further investigate the effect of the Eu valence on magnetic
ordering, we have performed band-structure calculations of the
generalized susceptibility $\chi(\textbf{q})$ for different valences
of Eu by varying the Fermi energy. In the $\chi(\textbf{q})$
calculation each small tetrahedron contribution (in \textbf{q}
space) was weighted by the Eu 5\textit{d} wavefunction components
which are predominantly responsible for coupling the Eu 4\textit{f}
moments via the RKKY mechanism. These calculations were performed
using the full-potential linearized augmented plane-wave (LAPW)
method with $R_{\textup{MT}}K_{\textup{max}}$~=~8 and
$R_{\textup{MT}}$~=~2.5, 2.2, and 2.2 a.u. for Eu, Rh, and As,
respectively. We used 405 \textit{k}-points in the irreducible
Brillouin zone for the self-consistent charge and 34061
\textit{k}-points in the whole reciprocal unit cell for the the
$\chi(\textbf{q})$ calculations. For the local density functional,
the Perdew-Wang 1992 functional\cite{perdew_92} was employed. The
convergence criterion for the total energy was 0.01 mRy/ cell.

\begin{figure}
\begin{center}
\includegraphics[clip, width=0.4\textwidth]{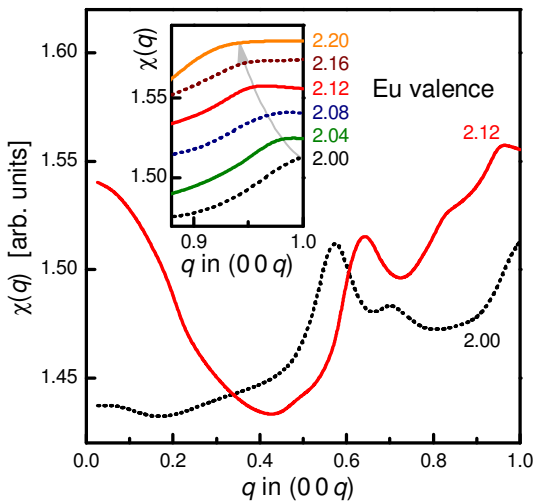}\\
\caption{(Color online) The generalized susceptibility
$\chi(\textbf{q})$ for different valence of Eu in EuRh$_2$As$_2$.
The inset shows an expanded view of $\chi(\textbf{q})$ close to the
Brillouin zone boundary at \textbf{q}~=~(0~0~1).} \label{chi}
\end{center}
\end{figure}

In Fig.~\ref{chi}, a distinct peak is evident in $\chi(\textbf{q})$
at \textbf{q}~=~(0~0~1) for divalent Eu and the peak moves
progressively to lower values of \textbf{q} as the valence is
increased (see inset to Fig.~\ref{chi} for details). We note the
presence of additional local maxima around \textbf{q}~=~(0~0~0.6)
and the zone center. Rather than attempt a detailed treatment of
RKKY matrix elements, we have calculated the total energy of the
virtual crystal with Eu$^{+2.1}$ for ferromagnetic (\textbf{q}~=~0)
and antiferromagnetic [\textbf{q}~=~(0~0~1)] ordering, and find that
the CM phase to be 9.0~meV lower in energy; thus eliminating
$\chi(\textbf{q}=0)$ peak from consideration. Therefore, band
structure calculations together with the relatively weak intensity
of the CM peaks suggest that the CM structure originates from a
minor phase associated with the divalent Eu ions and the ICM
structure from the major phase with an average valence of
$\sim$2.13, as inferred from the magnetization
measurement.\cite{singh_08}

We note that the temperature dependence of the CM peak is quite
different than the temperature dependence of the ICM peaks and the
value of the critical exponent is twice that of the ICM peak. For
the ICM peaks, the value of the critical exponent ($\beta$~=~0.32)
is close to that ($\beta$~=~0.36)\cite{kagawa_05} of the 3-D
classical Heisenberg model, typical for rare-earth elements in
intermetallic compounds.\cite{bruckel_01} As the temperature
dependence of surface magnetism
($\beta$~$\approx$~0.7)\cite{watson_96} can be quite different than
in the bulk, a surface bias of the minority phase could explain the
difference in the temperature dependences.

In summary, we have determined that below 46~K an ICM structure with
a temperature dependent propagation vector
$\boldsymbol{\tau}$~$\approx$~(0~0~0.9) coexists with a minor CM
structure for the magnetic order of Eu. The magnetic moments for
both the ICM and CM structures are within the tetragonal
\textbf{a-b} plane and are ferromagnetically aligned within this
plane. For the ICM structure, a spiral-like structure is most likely
with a turn angle of 162$^\circ$ between moments in adjacent Eu
planes. The existence of a spiral-like ICM structure down to the
lowest temperature indicates a weak in-plane anisotropy. For the CM
structure, magnetic moments in the adjacent Eu planes are
antiparallel aligned. Simultaneous occurrence of both the ICM and CM
structures and a different temperature dependence of the CM peak can
be explained with an additional minor phase and is consistent with
$\chi(\textbf{q})$ calculations, showing a strong sensitivity on the
Eu valence. Band structure calculations together with the observed
coexistence of CM and ICM phases indicate a delicate energy balance
between different magnetic configurations in EuRh$_2$As$_2$ and
makes this compound a promising candidate for studying the complex
interplay between changes in valence and magnetism as a function of
external parameters.

We thank D. S. Robinson for his help during experiments. The work at
the Ames Laboratory and at the MU-CAT sector was supported by the US
DOE under Contract No. DE-AC02-07CH11358. Use of the Advanced Photon
Source was supported by US DOE under Contract No. DE-AC02-06CH11357.

\bibliographystyle{apsrev}
\bibliography{eurh2as2_short}

\begin{thebibliography}{29}
\bibitem[{pv()}]{pv}
\bibinfo{note}{The propagation vectors (0~0~1) and (1~0~0) are equivalent for
  the space group $I4/mmm$. A magnetic structure or superstructure with such
  propagation vectors breaks the symmetry relations given by the body center.}
\bibitem[{sca()}]{scan}
\bibinfo{note}{($\xi$~0~8), (0~$\xi$~8) and ($\xi$~$\xi$~8) scans were
  performed with appropriate range and step size of $\xi$ and a counting time
  of 4~sec. for each data point.}
\bibitem[{att()}]{atten}
\bibinfo{note}{It was necessary to attenuate the incident beam by 75~\% to
  reduce beam heating, to properly characterize the temperature dependence of
  integrated intensity and to determine the transition temperature.
  Measurements shown in Figs.~\ref{escan}, \ref{tdep}, and \ref{qdep} are
  performed with this attenuation.}
\bibitem[{dl_()}]{dl_note}
\bibinfo{note}{The FWHM in scans along the (0~0~1) direction is dominated by
  the variation of the \textit{c} lattice parameter and is an order of
  magnitude larger than the intrinsic FWHM, as is evident from the linear
  extrapolation of FWHM to small $L$ values}.
\end{thebibliography}

\end{document}